\documentstyle[aps,12pt]{revtex}

\begin{document}
\title{Quantum secret sharing without entanglement}
\author{Guo-Ping Guo\thanks{%
Electronic address: harryguo@mail.ustc.edu.cn }, Guang-Can Guo\thanks{%
Electronic address: gcguo@ustc.edu.cn }}
\address{Key Laboratory of Quantum Information, University of Science and Technology\\
of China, Chinese Academy of Sciences, Hefei, Anhui, P.R.China, 230026}
\maketitle

\begin{abstract}
\baselineskip12ptAfter analysing the main quantum secret sharing protocol
based on the entanglement states, we propose an idea to directly encode the
qubit of quantum key distributions, and then present a quantum secret
sharing scheme where only product states are employed. As entanglement,
especially the inaccessable multi-entangled state, is not necessary in the
present quantum secret sharing protocol, it may be more applicable when the
number of the parties of secret sharing is large. Its theoretic efficiency
is also doubled to approach 100\%.

PACS number(s): 03.67.Hk, 89.70.+c
\end{abstract}

\baselineskip12pt

\section{Introduction}

Suppose the president of a bank, Alice, wants to give access to a vault to
two vice presidents, Bob and Charlie, who are not entirely trusted. Instead
of giving the combination to any one individual, it may be desirable to
distribute information in such a way that no vice president alone has any
knowledge of the combination, but both of them can jointly determine the
combination. Classical cryptography provides an answer which is known as
secret sharing\cite{f0}. Alice creates two coded messages and one of them is
sent to Bob and the other to Charlie. Each of the encrypted message contains
no information about her original message, but together they contain the
complete message.

However, either a fourth party or the dishonest member of the Bob-Charlie
pair gains access to both of Alice's transmissions can learn the contents of
her message in this classical procedure. Quantum secret sharing protocols
have been proposed to accomplish this work securely\cite{s1,s2,s3,s4.} where
multi-photon entanglement is employed. Recently, many kinds quantum secret
sharing with entanglement have been proposed\cite{s,s5,s6,s7}. As the
quantum key distribution and classical sharing protocol can be used
straightforward to accomplish secret sharing safely, the aim of the quantum
secret sharing protocols is to allow one to determine whether an
eavesdropper has been active during the secret sharing and reduce the
resources necessary to implement such multi-party secret sharing tasks\cite
{s1,s}. But the efficiency of those quantum secret sharing protocols using
entanglement can only approach 50\% in principle, the lack of the
multi-parties entanglement also holds their experimental implementation. The
efficiency of preparing even tripartite or four-partite entangled states is
very low\cite{du,du1}.

There is another thing that should be noted in this quantum secret sharing
scheme with GHZ states. It uses the correlation of the measurement value of
the three qubits of the GHZ state\cite{s1}: $\left\langle \sigma
_x^a\bigotimes \sigma _x^b\bigotimes \sigma _x^c\right\rangle =-\left\langle
\sigma _x^a\bigotimes \sigma _y^b\bigotimes \sigma _y^c\right\rangle
=-\left\langle \sigma _y^a\bigotimes \sigma _x^b\bigotimes \sigma
_y^c\right\rangle =-\left\langle \sigma _y^a\bigotimes \sigma _y^b\bigotimes
\sigma _x^c\right\rangle =1,$where $\sigma _j^i$ means the measurements
value $\pm 1$ of qubit $i$ along basis $j$ with $i=a,b,c$ and $j=x,y,z.$ So
only the cases that even of the three qubits are measured along the $y$
direction are kept to share secret in this scheme and its theory efficiency
is only $50\%.$

In the security analysis of the case that Bob eavesdrops both qubits $b$ and 
$c$, the authors assume that he measures them jointly in the $\left( \left|
00\right\rangle \pm \left| 11\right\rangle \right) /\sqrt{2}$or $\left(
\left| 00\right\rangle \pm i\left| 11\right\rangle \right) /\sqrt{2}$basis.
But we find if Bob measures qubits $b$ and $c$ separately (measures qubits $b
$ and $c$ in product states ), he can cheat and get Alice's secret without
being detected. For example, suppose Bob measures the qubits $b$ and $c$ in
the product basis $\sigma _y^b\bigotimes \sigma _y^c$, he can thus infer
Alice's value if she measures in the basis $\sigma _x^a$. Then Bob can send
a qubit in arbitrary state to Charlie. In the procedure of announcing basis,
Bob tries to announce his direction after Charlie and cheats his
announcement to let Alice keep the choose that she has measured his particle
in the basis $\sigma _x^a$. Different from the eavesdropping analysis in the
paper\cite{s1}, this eavesdropping will not introduce a higher than usual
failure rate. For example, if Charlie announces that he has chosen his
measurement in the basis $\sigma _x^c$, Bob can announce that his
measurement is also in the basis $\sigma _x^b$. Then if Alice also makes her
measurement in the basis $\sigma _x^a$, she will believe that there is
correlation between those three man's measurement value and keep this
measurement value as secret. But Bob has successfully eavesdrop this secret
value. 

Although Bob has changed the state of qubit sending to Charlie, he can cheat
again in the check procedure if Charlie announces his check bit before Bob
and then escapes detection. For example, when he had measured the qubits $b$
and $c$ in the product basis such as $\sigma _y^b\bigotimes \sigma _y^c$,
Bob can infer that if Alice measures her particle in the basis $\sigma _x^a$
will gets the value $+1$. When Charlie that his measurement is $-1$, Bob can
announces that his measurement is also $-1$. Then no error will occur in
this check procedure. Thus Bob successfully eavesdrops the secret without
being detected by Alice and Charlie. It means that if one party (say Bob)
can reveal his directions and values after the other party (say Charlie), he
is able to discover Alice's secret bit value, without any assistance from
Charlie, and escape being detected. The feature of this eavesdropping is
that Bob directly measures qubits $b$ and $c$ in product basis to learn
Alice's value. He cheats in the basis announcing procedure and check
procedure again. Bob needn't know Alice measurement direction, but he can
let Alice only keep the measurements he desired by announcing his direction
according to Charlie's measurement direction. Then this eavesdropping is
more powerful than those having been analyzed in the paper\cite{s1}.

In order to detect this kind of eavesdropping, Alice should randomly require
one party to reveal his directions before the other one, sometimes Bob
before Charlie, sometimes Charlie before Bob. And in the check procedure,
she also randomly requires one party to reveal his value before the other.
This random can ensure the security of this quantum secret sharing protocol
based on GHZ state correlation. It has also been pointed out in the paper%
\cite{s}, Alice and Bob can prevent this eavesdropping by releasing the
outcomes for the check bits before the announcement of their measurement
directions. We can see that the paper\cite{s} introduce the EPR state
correlations to split the secret. The efficiency of those quantum secret
sharing protocols with entanglement can only reach 50\% in principle.

In this paper, we propose the idea to directly encode the qubits of quantum
key distributions classically and present a quantum secret sharing scheme
employing product states to achieve the aim mentioned above. Its security is
based on the quantum no-cloning theory just as the BB84 quantum key
distribution. Comparing with the efficiency 50\% limiting for the existing
quantum secret sharing protocols with quantum entanglement, the present
scheme can be 100\% efficient in principle. It has been pointed out in the
paper\cite{s1,s} that the quantum key distribution can accomplish the task
of secret sharing, the resources necessary can be much smaller with our
present protocol although it is very simple and alike the BB84 quantum key
distribution scheme. Our emphasis is that the classically encoding the
qubits of quantum key distributions simply may be more economical than the
idea of the paper\cite{s} to use EPR\ states correlations to split secret.
This may suggests that the introduce of entanglement into the quantum secret
sharing may be unworthiness.

\section{Efficiency quantum secret sharing without entanglement}

Now we present a quantum secret sharing protocol where no entanglement is
employed. Its essence is to encode the sources of two BB84 key distribution
schemes. The particular process of this quantum secret sharing is as follows:

1. Alice creates two random $n$-bit string $L$ and $A$. For each bit of $L$
and $A$, she creates a two-qubit product state $\left| bc\right\rangle $ in
the basis $\oplus $ ( if the corresponding bit of $L$ is 0) or the basis $%
\otimes $ (if the bit of $L$ is 1) where the XOR result of the two qubits $b$
and $c$ equals to the bit value of the string $A$. The two-qubit vectors of
the basis $\oplus $ and $\otimes $ are $S1(\left| \alpha \beta \right\rangle
_{bc}^{\oplus })=\left\{ \left| 00\right\rangle _{bc},\left| 10\right\rangle
_{bc},\left| 01\right\rangle _{bc},\left| 11\right\rangle _{bc}\right\} $
and $S2(\left| \alpha ^{\prime }\beta ^{\prime }\right\rangle _{bc}^{\otimes
})=\left\{ \left| ++\right\rangle _{bc},\left| -+\right\rangle _{bc},\left|
+-\right\rangle _{bc},\left| --\right\rangle _{bc}\right\} $ respectively
where $\left| +\right\rangle =\frac 1{\sqrt{2}}(\left| 0\right\rangle
+\left| 1\right\rangle )$ and $\left| -\right\rangle =\frac 1{\sqrt{2}}%
(\left| 0\right\rangle -\left| 1\right\rangle )$ . Table 1 summarizes the
coding two-qubit states in the corresponding basis. For example, if the bits
of $L$ and $A$ are 1 and 0 respectively, then the two-qubit state Alice
prepared can be $\left| ++\right\rangle _{bc}$ or $\left| --\right\rangle
_{bc}$ each with 50\% probability. But she knows exactly which pair he
prepares.

2. Alice sends the resulting two strings $B$ and $C$ to Bob and Charlie
separately. It should be emphasized that every bit of the two strings $B$
and $C$ are corresponding and their XOR result equals to the corresponding
bit of the string $A$.

3. When both Bob and Charlie have announced the receiving of their strings $%
B $ and $C$, Alice announces the string $L$.

4. Bob and Charlie then measure each qubit of their string in the basis $%
\oplus $ or $\otimes $ according to the corresponding bit value of string $L$%
.

5. In the check procedure, Bob and Charlie are required to announce the
values of their check bits. If Alice finds too few of these values agree,
they abort this round of operation and restart from the first step.

6. The rest unchecked bits of strings $A$, $B$ and $C$ can be used as raw
keys for secret sharing. Alice's secret encrypted by her keys can only be
decrypted by Bob and Charlie when they cooperate with each other.

Obviously, the qubits $b$ and $c$ need not be sent out simultaneously in
this quantum secret sharing scheme. Alice can send them separately to Bob
and Charlie. They can hold their strings until Alice wants them to extract
out her information and announces her the string $L$. It is virtually a
combination of two modified BB84 quantum key distribution protocols, Alice
to Bob and Alice to Charlie, where the states Alice sent out have been
classically encoded so that her keys are shared in the strings $B$ and $C$.
Either of them can learn nothing about Alice's keys without the cooperation
of the other one. Furthermore, this secret sharing protocol is almost 100\%
efficient as all the keys can be used in the ideal case of no eavesdropping,
while the quantum secret sharing protocols with entanglement states\cite{s1}
can be at most 50\% efficient in principle. However, in this procedure of
making measurements after the announcement of bases, quantum memory is
required to store the qubits which has been shown available in the present
experimental technique\cite{gg}. Adjusting the measurements order in the
similar way the BB84 quantum key distribution protocol can also double its
efficiency of to nearly 100\%\cite{bb}. When no quantum memory is employed,
Bob and Charlie measure their qubits before Alice's announcement of basis in
our protocol, the efficiency of the present protocol falls to 50\%.

As the present scheme is more efficient than employing two BB 84 procedures,
one between Alice and Bob and the other between Alice and Charlie, which has
been briefly discussed in the references\cite{s1,s}, it allows the possible
eavesdropper, who obtains both of the particles that Alice sent, to gain
more information than the two-BB-84 scheme, when the qubits $b$ and $c$ are
sent simultaneously. Suppose an eavesdropper, Eve, obtains both of the
particles that Alice sent, and measures them in the Bell basis. Then when
the two qubits $b$ and $c$ are measured in the state $\left| \Phi
\right\rangle ^{-}=\frac 1{\sqrt{2}}(\left| 00\right\rangle -\left|
11\right\rangle )$, Eve can infers that the secret is $1$. And if they in
the state $\left| \Psi \right\rangle ^{+}=\frac 1{\sqrt{2}}(\left|
01\right\rangle -\left| 10\right\rangle )$, he knows that the secret is $0$.
But we can prevent this kind of information leakage causing by the jointed
measurements, either by sending the two qubit un-synchronously or by
encoding the two qubit as Table 2. Then in the same security degree, the
present protocol is double efficient as the two-BB-84 scheme. 

\section{Eavesdropping and generalization to the multi-party secret sharing}

In this quantum secret sharing scheme, only product states are employed and
one cannot get any information about the value of one qubit from measurement
on the other qubit. In fact it is exactly two independent BB84 quantum key
distribution protocols, which has been proved security generally\cite
{pp1,pp2}. The only difference is that the qubits Alice sent out have been
classical encoded. Obviously, any eavesdropping on individual quantum
channel for example Alice to Charlie will get nothing useful and will
unavoidably introduce errors.

Now suppose an adversary ( who could also be either Bob or Charlie)
eavesdrops the qubits $b$ and $c$ jointly. His object is to extract out
Alice's information without any assistance from Charlie in a way that cannot
be detected. But what ever he does, if Bob eavesdrops qubit $c$, Alice will
find some error in the check procedure although Bob could cheat again in
this check procedure. Then the security for the present quantum secret
sharing is guaranteed.

The generalization of this quantum secret sharing scheme to multi-parties is
straightforward.. For each bit of $L$ and $A$, Alice can creates an $N$%
-qubit product state $\left| b_1b_2...b_N\right\rangle $ in the basis $%
\oplus $ ( if the corresponding bit of $L$ is 0) or the basis $\otimes $ (if
the bit of $L$ is 1) where the XOR result of the $N$ qubits $b_1,b_2...b_N$
equals to the bit value of the string $A$. The two-qubit vectors of the
basis $\oplus $ and $\otimes $ are $S1(\left| \alpha _1\alpha _2...\alpha
_N\right\rangle _{12...N}^{\oplus })=\left\{ \left| 00...0\right\rangle
_{12...N},\left| 00...1\right\rangle _{12..N},...,\left| 11...1\right\rangle
_{12...N}\right\} $ and $S2(\left| \alpha _1^{\prime }\alpha _2^{\prime
}...\alpha _N^{\prime }\right\rangle _{12...N}^{\otimes })=\left\{ \left|
++...+\right\rangle _{12...N},\left| ++...-\right\rangle
_{12...N},...,\left| --...-\right\rangle _{12...N}\right\} $. Obviously, if
there are odd numbers $\left| 1\right\rangle $ in $\left| \alpha _1\alpha
_2...\alpha _N\right\rangle _{12...N}^{\oplus }$ ($\left| -\right\rangle $
in $\left| \alpha _1^{\prime }\alpha _2^{\prime }...\alpha _N^{\prime
}\right\rangle _{12...N}^{\otimes }$), the bit value of $A$ is 1. The case
of even numbers $\left| 1\right\rangle $ stands for 0. The following steps
are the same as the above two-party sharing case. As no entanglement,
especially inaccessible multi-parties entangled states, are required, it may
be more experimentally realizable than the original one\cite{s1}.

Finally, let us discuss the resources necessary to implement this quantum
secret sharing protocol. In the most obvious way of sharing secret, Alice
first must establish mutual keys among different pairs of parties, and then
use quantum cryptographic protocols to send each of the bit strings which
result from the classical procedure. While in our secret sharing protocol of
direct encoding the qubits of the two BB84 quantum key distribution, once
the key has been established, Alice needs to send only one string of
classical bits to either Bob or Charlie. Then in order to share a classical
bit(cbit) information between two parties, Bob and Charlie, quantum
cryptography as BB84 protocol ideally needs 2 qubits, 2 cbit and no ebit on
average. The quantum secret sharing with EPR states needs 4 qubits, 1cbit
and 2ebits. While our protocol with product states needs only 2 qubits,
1cbit and no ebit. In general, the more parts into which the secret is
split, the greater the difference between the number of classical bits which
be sent in those two ways. We see the direct encoding to the source qubits
is able to act as a substitute for transmitted random bits. Furthermore, the
efficiency of the existing entangled-state scheme can only reach 50\% in
principle while the present secret sharing protocol can be 100\% efficient.

\section{Conclusion}

In this paper, we analyze the main quantum secret sharing protocol based on
the GHZ-state and propose a quantum secret sharing scheme where only product
states are employed. As entanglement, especially the inaccessible
multi-party entangled state, is not necessary in the present quantum secret
sharing protocol, it may be more applicable when the number of the parties
of secret sharing is large. Its theoretic efficiency is also doubled to
approach 100\%.

Although those quantum secret sharing with entanglement can be a
demonstrator of concepts of various entanglement applications, the inherent
efficiency limiting of these protocols and the difference for the
preparation of multi-party entanglement may suggest that the employing of
the quantum entanglement in quantum secret sharing be unworthiness in some
cases. After all the practicality is an important pursuit of the quantum
information theory.

This work was funded by National Fundamental Research Program(2001CB309300),
National Natural Science Foundation of China, the Innovation funds from
Chinese Academy of Sciences, and also by the outstanding Ph. D thesis award
and the CAS's talented scientist award entitled to Luming Duan.

\end{document}